\documentclass[conference]{IEEEtran}
\IEEEoverridecommandlockouts

\usepackage{orcidlink} 

\usepackage{cite}
\usepackage{amsmath,amssymb,amsfonts}
\usepackage{algorithmic}
\usepackage{graphicx}
\usepackage{textcomp}
\usepackage{xcolor}
\usepackage[numbers,square,sort&compress]{natbib}
\hypersetup{
    hidelinks
}
\usepackage{cleveref}
\def\BibTeX{{\rm B\kern-.05em{\sc i\kern-.025em b}\kern-.08em
    T\kern-.1667em\lower.7ex\hbox{E}\kern-.125emX}}
    
\usepackage{listings}
\usepackage{float}

\usepackage{tabularx}

\usepackage{enumitem}
\usepackage{textcomp}

\usepackage{listings}    
\usepackage{xcolor}      
\usepackage{subcaption} 

\newcolumntype{Y}{>{\centering\arraybackslash}X}

\newfloat{lstfloat}{htbp}{lop}
\floatname{lstfloat}{Listing}

\definecolor{background}{HTML}{EEEEEE}
\definecolor{json_string}{RGB}{20,105,176}
\colorlet{json_default}{magenta!60!black}
\definecolor{json_number}{RGB}{0,0,0} 
\definecolor{line_color}{RGB}{220, 0, 40}

\definecolor{deepblue}{rgb}{0,0,0.5}
\definecolor{deepred}{rgb}{0.6,0,0}
\definecolor{deepgreen}{rgb}{0,0.5,0}
\definecolor{lightgray}{rgb}{0.95,0.95,0.95}
\definecolor{red}{rgb}{1,0,0}
\definecolor{sand}{rgb}{0.65, 0.48, 0.02}
\definecolor{lightblue}{rgb}{0.02, 0.58, 0.65}

\DeclareFixedFont{\ttb}{T1}{txtt}{bx}{n}{12} 
\DeclareFixedFont{\ttm}{T1}{txtt}{m}{n}{12}  

\newcommand\pythonstyle{\lstset{
    language=Python,
    basicstyle=\small\ttfamily,
    morekeywords={self},                   
    keywordstyle=\small\color{deepblue},
    emph={__init__,np,sklearn,dille,nbconvert,linear_model,LogisticRegression,numpy,pandas,torch,pd,preprocessing,StandardScaler,DataFrame,random,GaussianNB,naive_bayes,svm,SVC}, 
    emphstyle=\small\color{deepred}, 
    emph={[2]model,X,y,data,columns,df,scaler,scaled_data,scaled_df,gnb,svc},                  
    emphstyle={[2]\small\color{lightblue}}, 
    emph={[3]array,fit,nan,rand,fit_transform,transform,read_csv},                  
    emphstyle={[3]\small\color{sand}}, 
    stringstyle=\color{deepgreen},
    frame=tb,                              
    showstringspaces=false,
    numbers=left,                          
    numberstyle=\scriptsize\color{red},    
    stepnumber=1,                          
    numbersep=4pt,                         
    backgroundcolor=\color{lightgray},     
    xleftmargin=1.5em,                       
    captionpos=b
}}

\lstnewenvironment{python}[1][]
{
\pythonstyle
\lstset{#1}
}
{}

\DeclareRobustCommand\pythoninline[1]{{\pythonstyle\lstinline!#1!}}

\newcommand\jsonstyle{\lstset{
    language={},
    basicstyle=\small\ttfamily,
    frame=tb,
    showstringspaces=false,
    numbers=left,
    numberstyle=\scriptsize\color{red},
    stepnumber=1,
    numbersep=4pt,
    backgroundcolor=\color{lightgray},
    xleftmargin=1.5em,
    captionpos=b,
    morekeywords={true,false,null},
    keywordstyle=\small\color{deepred},
    morestring=[b]",
    stringstyle=\color{deepgreen}, 
    moredelim=**[is][\color{deepblue}]{@key}{@}, 
    moredelim=**[is][\color{deepgreen}]{@val}{@}, 
    literate=
       {\{}{{{\color{black}\{}}}1
       {\}}{{{\color{black}\}}}}1
       {[}{{{\color{black}[}}}1
       {]}{{{\color{black}]}}}1
       {:}{{{\color{black}:}}}1
       {,}{{{\color{black},}}}1
}}

\lstnewenvironment{json}[1][]
{
  \jsonstyle
  \lstset{#1}
}
{}

\DeclareRobustCommand\jsoninline[1]{{\jsonstyle\lstinline!#1!}}

\newcommand{\Description}[1]{}

\newif\ifblindreview

\blindreviewfalse  

\ifblindreview
  \newcommand{\anon}[1]{{\color{orange}ANONYMIZED}}
\else
  \newcommand{\anon}[1]{#1}
  
\fi

\begin{document}

\title{Data-aware Static Analysis: Improving Detection of Semantic Faults in Machine Learning Code Using Data Characteristics
\thanks{This work was partially supported by the Wallenberg AI, Autono\-mous Systems and Software Program (WASP) funded by the Knut and Alice Wallenberg Foundation, and was in collaboration with Software Center Project 61 and Vinnova CoDig competence center.}}

\ifblindreview
\author{\IEEEauthorblockN{\textit{Anonymous Authors}}}
\else
\author{\IEEEauthorblockN{Willem Meijer~\orcidlink{0000-0001-8482-3917}}
\IEEEauthorblockA{\textit{Linköping University}\\
Linköping, Sweden \\
willem.meijer@liu.se
}
\and
\IEEEauthorblockN{Kristian Sandahl~\orcidlink{0000-0002-3052-5604}}
\IEEEauthorblockA{\textit{Linköping University}\\
Linköping, Sweden \\
kristian.sandahl@liu.se
}
\and
\IEEEauthorblockN{Dániel Varró~\orcidlink{0000-0002-8790-252X}}
\IEEEauthorblockA{\textit{Linköping University}\\
Linköping, Sweden \\
daniel.varro@liu.se
}
}
\fi

\maketitle

\begin{abstract}
Semantic faults specific to the use of machine learning models are a common problem for machine learning developers, causing suboptimal predictions, high computational cost, or incorrect outputs.
For example, one may erroneously use unscaled data to train a scale-sensitive model.
Machine learning developers detect these faults after training their models and manually analyzing the results, making it an inefficient process.
We propose a novel data-aware static analysis approach to detect semantic faults in machine learning code, allowing developers to reveal these bugs while writing code instead of after training the model.
Our approach uses combined data and control flow analysis, and API contracts, enabling data-aware reasoning about machine learning code at a high level of abstraction.
We highlight the potential of our solution by analyzing a sample of real-world machine learning notebooks, finding that we can detect faults that require a data-aware approach.

\end{abstract}


\section{Introduction}

With the increase in adoption of machine learning-powered systems~\citep{oecdbcginsead_adoption_2025, arroyabe_analyzing_2024}, the need for quality assurance techniques has grown as well~\citep{shivashankar_maintainability_2022, santhanam_quality_2020, bogner_characterizing_2021, cote_quality_2024}.
Various attempts have been made to enrich this field, addressing architectural concerns~\citep{nazir_architecting_2024, bucaioni_checklist_2025}, best practices for machine learning development~\citep{serban_software_2024}, data validation~\citep{whang_data_2023, kumar_opportunities_2024, albelali_testing_2025}, or machine learning system testing~\citep{albelali_testing_2025}.
However, tool support for effectively building machine learning models remains limited, as machine learning developers still commonly struggle with \textit{semantic faults} specific to machine learning models~\citep{wang_why_2025, humbatova_taxonomy_2020, de_santana_bug_2024, khairunnesa_what_2023}.
For example, one may use unscaled data to train a scale-sensitive model.
Such semantic faults result in poor or suboptimal prediction performance, excessive compute resource usage, or incorrect outputs.

In their current practice, machine learning developers detect semantic faults in machine learning models after training the model, manually investigating the training results, determining if suboptimal results are due to a bug or just the wrong choice of model, locating the bug in the pipeline and/or data, fixing the bug, and re-training the model; after which the cycle repeats.
As a consequence, debugging semantic faults in machine learning is currently a highly time- and resource-consuming process~\citep{lai_comparative_2024, morovati_bug_2024}.

Existing solutions may identify incorrect hyperparameters and certain training issues~\citep{gao_refty_2022, reimann_safe-ds_2023, shivashankar_mlscent_2025, ahmed_design_2023}, but have fundamental limitations: they do not consider real data at all~\citep{shivashankar_mlscent_2025, ahmed_design_2023, turcotte_fault_2025}, or only consider structural properties~\citep{gao_refty_2022, reimann_safe-ds_2023}, not semantic properties like distributions or correlations.
Further, these solutions only provide feedback at runtime instead of during programming~\citep{gao_refty_2022, reimann_safe-ds_2023, ahmed_design_2023, turcotte_expressing_2025} or rely on impractical assumptions to identify faults~\citep{shivashankar_mlscent_2025, turcotte_fault_2025}.

\textit{Our work proposes a novel data-aware approach to detect semantic faults statically in machine learning code.}
Static detection allows machine learning developers to detect faults while writing their code, before training their models, thus reducing the manual investigation of the results.
We leverage two well-known techniques for program analysis: 1)~combined data and control flow analysis, and 2)~contracts for application programming interfaces (APIs).
This enables data-aware reasoning about machine learning code at a high level of abstraction after extracting complete machine learning pipelines to reveal what data is loaded, how it is preprocessed, and what models are trained with it.
In turn, we define contracts for machine learning APIs that capture the semantic prerequisites of machine learning models (like data distributions), and evaluate if these are met by analyzing the extracted pipeline on real data.

We show our solution's potential impact with a preliminary analysis in the context of the popular \pythoninline{sklearn} library.
Even with a limited number of contracts evaluated on a small number of randomly sampled real-world machine learning notebooks, we can already detect semantic faults that require a data-aware approach.

\newcommand{\ftnSVC}{\footnote{SVC documentation: \href{https://scikit-learn.org/stable/modules/generated/sklearn.svm.SVC.html}{https://scikit-learn.org/stable/modules/generated/sklearn.}\\\href{https://scikit-learn.org/stable/modules/generated/sklearn.svm.SVC.html}{svm.SVC.html}}}

\newcommand{\ftnSS}{\footnote{StandardScaler documentation: \href{https://scikit-learn.org/stable/modules/generated/sklearn.preprocessing.StandardScaler.html}{https://scikit-learn.org/stable/modules/generated/}\\\href{https://scikit-learn.org/stable/modules/generated/sklearn.preprocessing.StandardScaler.html}{sklearn.preprocessing.StandardScaler.html}}}

\section{Running Example}

\begin{figure}[!t]
\begin{tabular}{cc} 
        \includegraphics[width=0.475\linewidth]{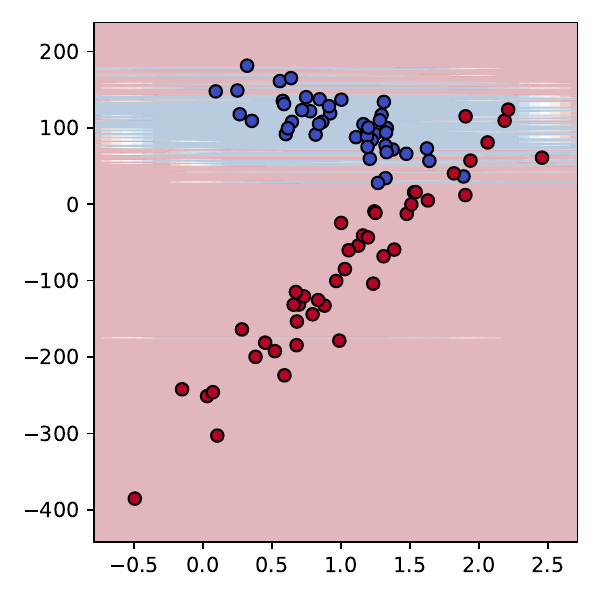} 
        &  \includegraphics[width=0.475\linewidth]{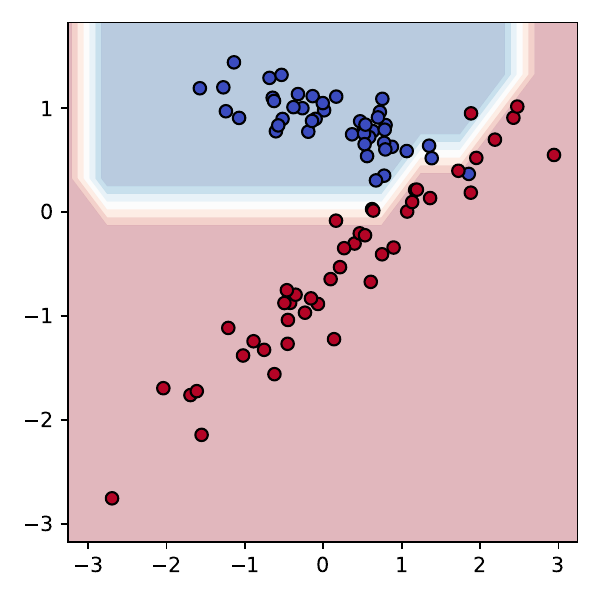}
\end{tabular}
\caption{Decision boundary plots of a \pythoninline{SVC} fitted on unequally scaled data (left) versus equally scaled data (right)}
\Description{Decision boundary plots of a \pythoninline{SVC} fitted on unequally scaled data (left) versus equally scaled data (right)}
\label{fig:running-example-data}
\end{figure}

To illustrate the type of faults we aim to detect, we include \autoref{lst:example}~\citep{meijer_contract-based_2024}, implementing a simple machine learning pipeline for binary classification.
It imports some packages on lines 1 and 2, loads a dataset \pythoninline{df} on line 3, constructs a support vector machine\ftnSVC~\pythoninline{svc} on line 4, and trains it on the dataset on line 5 using the \pythoninline{fit} method.
Seemingly, this code is syntactically correct and runs without exceptions.
However, it can still fail semantically, as it does not account for the support vector machine's prerequisite that the scale of each feature in the input data needs to be roughly equal.
(Note that this is not mentioned in the official documentation.)
This issue is highlighted in \autoref{fig:running-example-data}, where the model trained on unscaled data clearly overfits, as indicated by its complex decision boundary.
By scaling the input data, for example, by using \pythoninline{sklearn}'s \pythoninline{StandardScaler},\ftnSS~the same code creates an easier-to-interpret, more robust, and better-performing result.
Without considering both the data and the code, one cannot know if this transformation is required or if the dataset was already scaled correctly when it was loaded.




\begin{lstfloat}[!t]
\begin{python}[caption={Example of a support vector classifier (\pythoninline{SVC}) containing a semantic fault when the data is not equally scaled~\cite{meijer_contract-based_2024}\Description{Example of a support vector classifier (\pythoninline{SVC}) containing a semantic fault when the data is not equally scaled~\cite{meijer_contract-based_2024}}}, label={lst:example}]
import pandas as pd
from sklearn.svm import SVC
df = pd.read_csv('my_dataset.csv')
svc = SVC(kernel='rbf', gamma='auto')
svc.fit(df[['x1', 'x2']], df['y'])
\end{python}
\end{lstfloat}

\section{State of the Art in Semantic Fault Detection}
    Some methods exist to detect semantic faults in machine learning systems, which commonly separate data from code ~\citep{whang_data_2023, kumar_opportunities_2024}.
    For example, \citet{hynes_data_2017} explore data cleaning and validation by writing test cases for dataset properties such as minima or averages, verifying data points using domain knowledge.
    Similarly, \citet{schelter_jenga_2021} evaluate a pipeline's robustness by intentionally injecting data faults, like missing data, into the dataset.
    Although these purely data-centric methods can find data issues and test the robustness of machine learning pipelines, they are unable to detect semantic faults specific to particular machine learning models.
    
    \autoref{tab:related-work} highlights state-of-the-art solutions that detect semantic faults in machine learning~\citep{gao_refty_2022, reimann_safe-ds_2023, shivashankar_mlscent_2025, ahmed_design_2023} and statistics code~\citep{turcotte_expressing_2025, turcotte_fault_2025}.
    We distinguish between data-aware and data-agonistic techniques, illustrating whether the analysis actively uses properties of training or testing data.
    To highlight when solutions identify faults, we further distinguish between static and runtime analysis, respectively identifying faults while writing and when running the code.

\begin{table}[!t]
    \caption{Overview of the state of the art, highlighting whether they are data-aware/agnostic and use static/runtime analysis}
    \label{tab:related-work}
    \centering
    \begin{tabularx}{\columnwidth}{|c|Y|Y|}
        \hline
         & \textbf{Static} & \textbf{Runtime} \\
        \hline\hline
        \textbf{Data-aware} & \textit{Our solution} & \citep{reimann_safe-ds_2023, gao_refty_2022, turcotte_expressing_2025} \\
        \textbf{Data-agnostic} & \citep{shivashankar_mlscent_2025, turcotte_fault_2025} & \citep{reimann_safe-ds_2023, gao_refty_2022, ahmed_design_2023} \\
        \hline
    \end{tabularx}
\end{table}

    \paragraph{Data-agnostic Techniques}
        Various solutions identify semantic faults in code without using data~\citep{ahmed_design_2023, gao_refty_2022, reimann_safe-ds_2023, shivashankar_mlscent_2025, turcotte_fault_2025}.
        Like us, they apply contract-like solutions to enforce API constraints.
        Most commonly, these are hyperparameter constraints that detect missing or incompatible hyperparameters~\citep{ahmed_design_2023, gao_refty_2022, shivashankar_mlscent_2025, reimann_safe-ds_2023}.
        For example, the \pythoninline{kernel} in \pythoninline{sklearn}'s support vector classifier has five valid options.
        Two solutions extend this~\citep{shivashankar_mlscent_2025, turcotte_fault_2025} by statically testing for data-related issues using heuristics instead of real data.
        For example, you could use whether a dataset was visualized as an indicator of class imbalance (a common cause of bias in models) or if statistical assumptions are visually verified.
        While data visualization can be used to indicate these issues, it does not guarantee they are properly addressed.
        \citet{ahmed_design_2023} detect training issues in deep learning code using runtime analysis.
        They identify problems such as slow convergence in deep learning models by observing training results.
    
    \paragraph{Data-aware Techniques}
        Some solutions exist that explicitly include data into their analysis~\citep{gao_refty_2022, reimann_safe-ds_2023, turcotte_expressing_2025}.
        \citet{gao_refty_2022} include data shapes and formats to identify whether the dataset structure is compatible with the used deep learning model.
        \citet{reimann_safe-ds_2023} developed a domain-specific language for machine learning pipelines, complemented by rules that verify whether accessed column names exist or whether a model was trained before it was used for prediction.
        Although these methods detect structural issues, they cannot detect semantic faults specific to machine learning models, such as incorrect data distributions.
        Although \citet{turcotte_expressing_2025} transcend these methods by injecting statistical tests that verify model assumptions (e.g., feature independence), they include these tests at runtime.

\paragraph{Novelty}
    Up to our best knowledge, our approach is the first to provide a data-aware static analysis technique to reveal semantic faults related to the incorrect use of specific machine learning models. Moreover, our approach covers a broader range of semantic faults, including distribution and correlation faults, in addition to inappropriate hyperparameters and structural issues. 



\newcommand{\ftnPylint}{\footnote{Pylint documentation: \href{https://pypi.org/project/pylint/}{https://pypi.org/project/pylint/}}}
\newcommand{\ftnPyAst}{\footnote{Python AST documentation: \href{https://docs.python.org/3/library/ast.html}{https://docs.python.org/3/library/ast.html}}}

\newcommand{\ftnPT}{\footnote{Pytorch documentation: \href{https://pypi.org/project/torch/}{https://pypi.org/project/torch/}}}

\newcommand{\ftnNorm}{\footnote{L2 norm documentation: \href{https://docs.scipy.org/doc/scipy/reference/generated/scipy.linalg.norm.html}{https://docs.scipy.org/doc/scipy/reference/generated/scipy.}\\\href{https://docs.scipy.org/doc/scipy/reference/generated/scipy.linalg.norm.html}{linalg.norm.html}}}

\section{Analysis Procedure}
\label{sec:implementation}
    To improve semantic fault detection in machine learning code, we propose an approach that natively exploits training and testing data.
    At a high level, our approach uses combined data and control flow analysis to capture how data is used in the code, including where it is loaded, how it is preprocessed, and which models are trained on it.
    Using this representation, we can identify the prerequisites of the used machine learning models as contracts, and check whether the real data, in combination with the preprocessing steps, are compliant with those contracts.
    We carry out our analysis in three main steps: 1) abstract syntax tree (AST) transformation, 2) data and control flow extraction, and 3) contract evaluation.
    Respectively, these steps rewrite Python code to a simplified and standardized format, transform this into a directed acyclic graph representation, and evaluate whether the machine learning APIs are used correctly.
1    
    

\subsection{AST Transformation}
    To simplify the analysis procedure described in Section~\ref{sec:data-control-flow}, we first transform the AST of Python scripts to simplify their coding style.
    For brevity, we refer to our experimental code for concrete examples~\citep{meijer_experiment_2025}.
    Firstly, we unfold self-defined functions such that their code is executed at the module level (i.e., file level).
    Developers commonly define their own methods to, e.g., create a common preprocessing pipeline.
    We continue by unwrapping well-defined loops: loops that iterate over a hard-coded series of elements.
    Developers commonly use these during, e.g., model training, to compare a pre-specified list of models in an attempt to identify the best one.
    We then continue by extracting nested statements and splitting long chained statements, two programming anti-patterns that are overwhelmingly present in machine learning notebooks in our dataset~\citep{wang_why_2025}.
    Finally, we resolve all aliases to their full names and prefix any imported name with its full library reference.

\begin{figure}[!t]
    \includegraphics[width=0.85\linewidth]{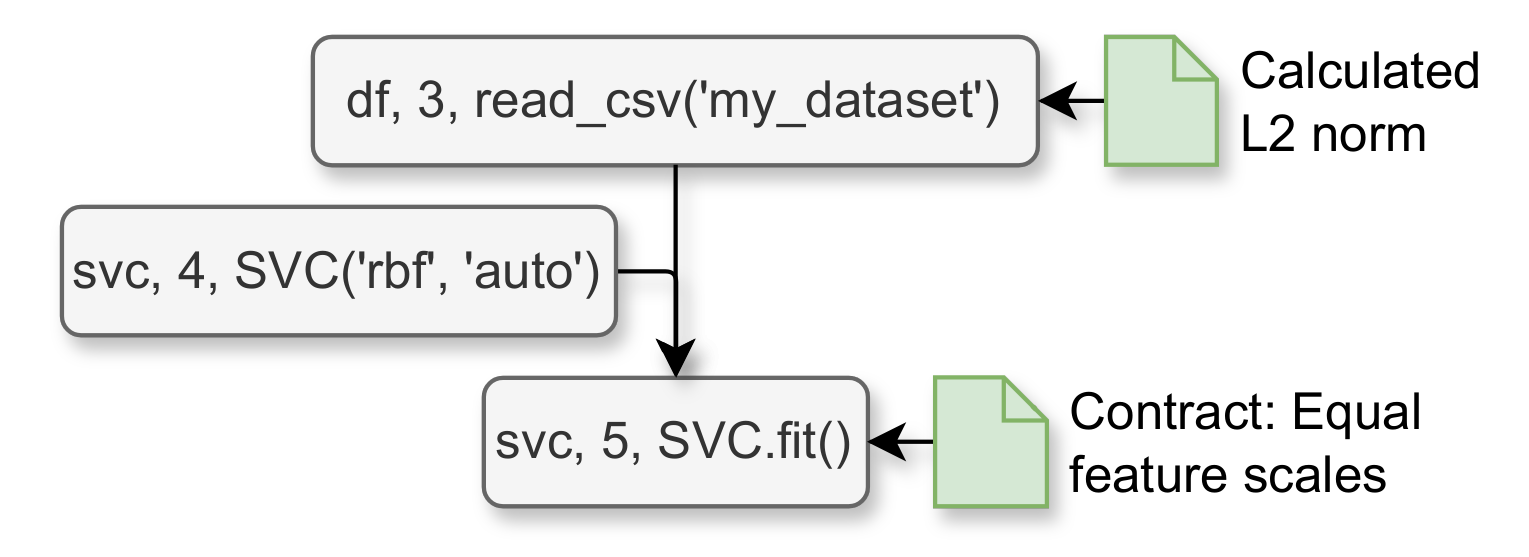}
    \caption{Visual representation of the running example in \autoref{lst:example}, highlighting the API contract and calculated data properties in green at the right}
    \Description{Visual representation of the running example in \autoref{lst:example}, highlighting the API contract and calculated data properties in green at the right}
    \label{fig:bpmn-implementation}
\end{figure}

\subsection{Data and Control Flow Extraction}\label{sec:data-control-flow}
    Next, we use preprocessed ASTs to extract data and control flow representations.
    Conceptually, this allows us to represent our running example of \autoref{lst:example} as the directed acyclic graph shown in \autoref{fig:bpmn-implementation} (ignoring the contract and data property nodes).
    This representation is necessary because the machine learning scripts in our dataset commonly define multiple pipelines to compare their performance, complicating the identification of which preprocessing steps were applied to which datasets.
    Because our AST transformations simplify the code, the pipeline's data and control flow can be extracted by iterating through all function calls and assignments at the module level.
    The notebooks in our dataset seldom contain custom classes, allowing us to omit internal state in custom objects (this is not true for deep learning models implemented using, e.g,~\pythoninline{torch},\ftnPT which are subject to future work).
    By iterating through assignments, we extract name-line-value triples that represent the value of a name at some location in the script.
    For example, given line 3 in our running example, \pythoninline{df = read\_csv('my_dataset.csv')}, we extract the triple $\langle$~\pythoninline{df},~3,~\pythoninline{read\_csv('my_dataset.csv')}~$\rangle$.
    Consequently, any future use of \pythoninline{df} (e.g., to train a model) can be traced to its most recent assignment, making it possible to traverse backwards through the graph.

\begin{lstfloat}[!t]
\begin{json}[caption={Example of an API contract, stating the targeted API, constructor prerequisites, description, and the method that evaluates the contract\Description{Example of an API contract, stating the targeted API, constructor prerequisites, description, and the method that evaluates the contract}}, label={lst:contract-example}]
"api": "sklearn.svm.SVC",
"endpoint": "fit",
"with_construct_param": { "kernel": "rbf" },
"rule_name": "Data Must Have Equal Scales",
"importance": "Critical",
"description": "The data used to fit the SVC
  must be equally scaled because ...",
"eval_func": "evaluators.max_l2_norm_ratio",
"variables": { "threshold": 10 }
\end{json}
\end{lstfloat}

\subsection{Contract Evaluation}

\subsubsection{Machine Learning API Contracts}
    We define API contracts for machine learning API endpoints.
    API contracts are well-known concepts in program analysis, defining how APIs should be used.
    A machine learning contract, with an example shown in \autoref{lst:contract-example}, has four basic components: 1)~the related API, 2)~the parameter prerequisites for that contract to be relevant (some contracts are only relevant when a specific hyperparameter is used), 3)~a description of the contract, and 4)~a reference to an \textit{evaluator method} that evaluates the contract.
    Currently, we define three contract categories: a)~hyperparameter restrictions (enforcing the use of supported individual and combinations of hyperparameters), b)~parameterized model requirements (testing whether the training and testing data have similar enough distributions), and c)~data property requirements (enforcing data distribution properties, like feature ranges or cardinality, and statistical properties, like the L2 norm,\ftnNorm~which we use to evaluate our running example).

\subsubsection{Data Property Calculation}\label{sec:reverse-pipeline-traversal}
    To evaluate our contracts, we first iterate through all method calls in the AST and test whether related contracts exist.
    Then, using the variable names specified in this call, we can identify the most recent assignment to these variables through our data and control flow representation.
    For example, the \pythoninline{SVC.fit} call specifies its input dataset \pythoninline{df} and the model instance \pythoninline{svc}.
    This information can then be used to identify what contracts correspond to the model's hyperparameters.
    Moreover, we can identify the last assignment of the used dataset.
    In the running example, this would lead to the \pythoninline{read_csv} method.
    This is ideal as \pythoninline{read_csv} concretely specifies the dataset path, meaning we can load (a random sample of) data into memory and use that to calculate the evaluated data property and verify if it satisfies our contract.

\subsubsection{Guarantees for Preprocessing APIs}
\begin{figure}[!t]
    \centering
    \includegraphics[width=0.90\linewidth]{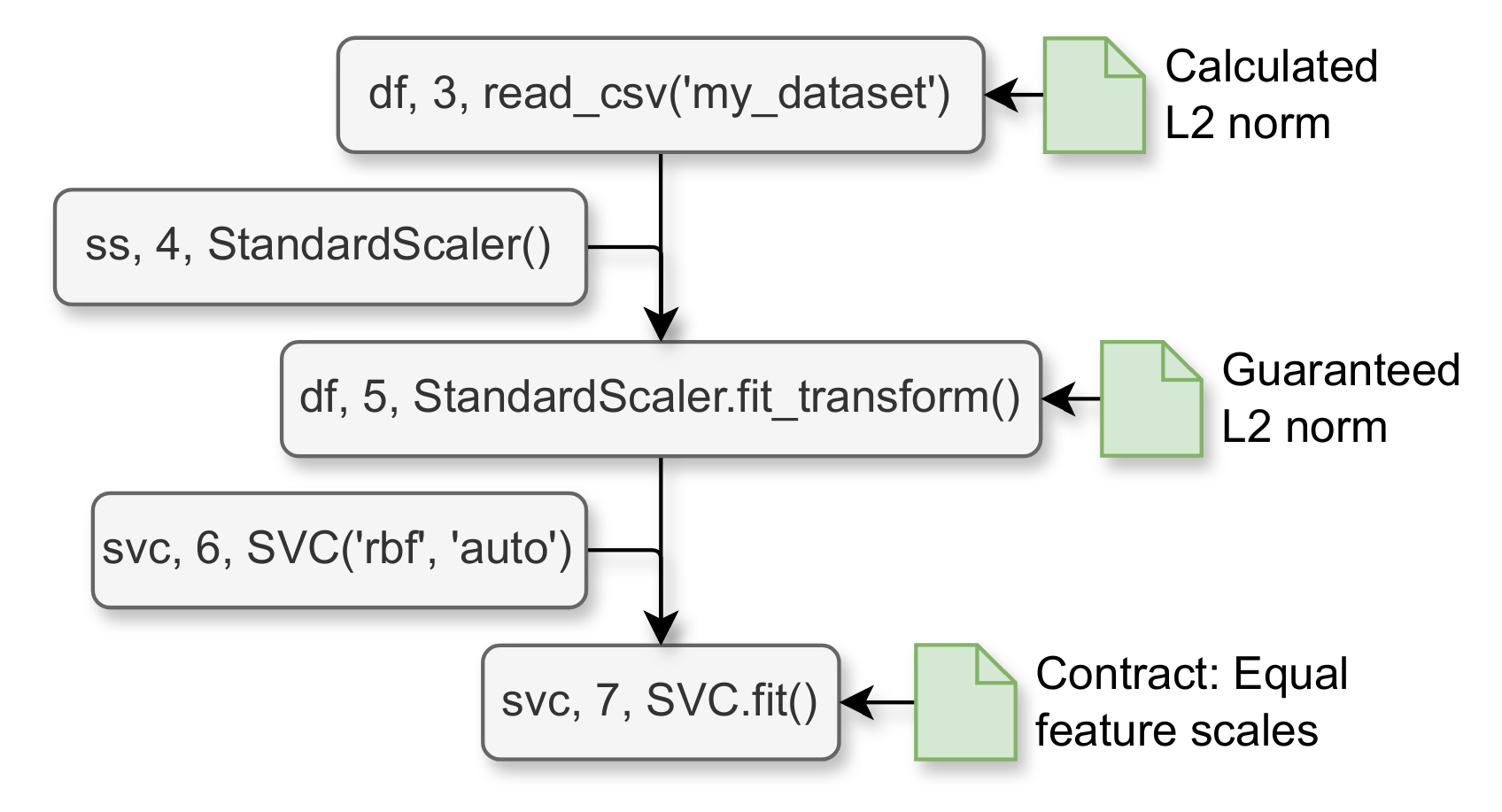}
    \caption{Extension of Figure~\ref{fig:bpmn-implementation} where the dataset is transformed using {\pythoninline{StandardScaler.fit_transform}}, showing its corresponding L2 norm guarantee in green at the right}
    \Description{Extension of Figure~\ref{fig:bpmn-implementation} where the dataset is transformed using {\pythoninline{StandardScaler.fit_transform}}, showing its corresponding L2 norm guarantee in green at the right}
    \label{fig:bpmn-implementation-extended}
\end{figure}

    Unlike in the running example, it is very common that preprocessing is applied to the data before it is used to train a machine learning model.
    These perform transformations such as imputation, scaling, or data resampling.
    An example is shown in Figure~\ref{fig:bpmn-implementation-extended}, which extends our running example by adding the \pythoninline{StandardScaler} to transform the dataset.
    Because our solution analyzes code statically, we cannot directly calculate output data properties, as we do not have access to runtime data.
    Therefore, we implement guarantees for preprocessing APIs that capture the assumptions that can be made about the data after a preprocessing step.
    The specification of these guarantees is almost identical to the contract shown in \autoref{lst:contract-example}.
    However, in addition to an API, it also specifies a data property.
    
    A simple example of an API guarantee is the data's mean after using \pythoninline{StandardScaler.fit_transform}, which calculates the dataset's mean and standard deviation for each feature and uses them to apply the $z$-transform to each feature.
    By definition, this allows us to assume that each feature in the output data has a mean of zero without running any additional computation.
    Consequently, we can write a guarantee stating that the mean of each feature in the output of \pythoninline{StandardScaler.fit_transform} is always zero.

    Because some guaranteed properties are not as well-defined, they require some additional calculation.
    An example of this is \pythoninline{sklearn}'s \pythoninline{StandardScaler.fit_transform} when calculating the L2 norm.
    In that case, we cannot assume it has some constant value.
    However, because the formula to calculate the new L2 norm is well-defined, we can write a custom method that extracts the relevant data properties of the method's input data (again, using reverse pipeline traversal as elaborated in Section~\ref{sec:reverse-pipeline-traversal}), calculate the updated L2 norm, and then use that to evaluate the contract.

\newcommand{\ftnJP}{\footnote{Jupyter documentation: \href{https://docs.jupyter.org/en/latest/}{https://docs.jupyter.org/en/latest/}}}

\newcommand{\ftnNBC}{\footnote{nbconvert documentation: \href{https://nbconvert.readthedocs.io/en/latest/}{https://nbconvert.readthedocs.io/en/latest/}}}

\newcommand{\ftnMGC}{\footnote{Magic command documentation: \href{https://ipython.readthedocs.io/en/stable/interactive/magics.html}{https://ipython.readthedocs.io/en/stable/interactive}\\\href{https://ipython.readthedocs.io/en/stable/interactive/magics.html}{/magics.html}}}

\section{Preliminary Analysis}
    An early prototype of our framework and our experimental code is available on GitHub~\citep{meijer_experiment_2025, meijer_implementation_2025}.~
    As of now (\pythoninline{dille} version \pythoninline{0.8.0}), we have implemented 189 contracts for 52 API endpoints and 60 guarantees for 11 API endpoints.
    We have focused on \pythoninline{pandas} and \pythoninline{sklearn} endpoints, as these are the most common non-deep learning data management and machine learning libraries~\citep{wang_why_2025}.
    
    To evaluate the potential impact of our solution, we performed a preliminary analysis.
    We leverage the dataset shared by \citet{wang_why_2025}, containing roughly $1.2$ million Jupyter notebooks.\ftnJP~
    Notebooks provide an interactive interface for Python scripts (among others), making them fundamental in data science due to their ability to give real-time feedback to developers~\citep{wang_better_2020}.
    We randomly sampled 21 1) Kaggle notebooks 2) written in syntactically valid Python 3) that use any \pythoninline{sklearn} API supported by our tool, and 4) have available \pythoninline{CSV} datasets.
    Because our solution does not natively support notebooks, we transformed these to Python scripts using \pythoninline{nbconvert}.\ftnNBC

    Using our solution, we detected semantic faults in 5 out of 21 scripts.
    These scripts contained 17 unique contract violations, related to six unique contracts.
    Three of these contracts were data-specific, responsible for seven violations, and the other three contracts were hyperparameter constraints.
    Notably, all data-specific violations occurred in the same script, being marked for redundant feature imputation five times, once for using a cardinality-sensitive algorithm with a high-cardinality feature, and once for using unequal data ranges to train a range-sensitive algorithm.  
    After investigation, we concluded that five out of seven marked violations were correct and two were incorrect.
    Both of these misclassifications marked redundant imputation steps, which were triggered due to the absence of missing data in the sample we took from the complete dataset, as it was quite rare.
    Consequently, if the complete dataset was used, these issues would not have been marked.

    Even with this small empirical sample, we were already able to use our prototype to identify a series of semantic faults in machine learning code.
    Because almost half of these were data-specific, this highlights the importance and potential of a data-informed approach when evaluating machine learning software.

\section{Conclusion}

We propose a novel data-aware static analysis approach to detect semantic faults related to the use of specific machine learning models.
Developers can use this method to detect semantic faults while writing code instead of after training the model, thus reducing the time needed to resolve faults.
Our approach uses combined data and control flow analysis, and API contracts, enabling data-aware reasoning about machine learning code at a high level of abstraction.
We highlight the potential of our solution by analyzing some real-world scripts that use the popular \pythoninline{sklearn} library, finding that we can detect faults that require a data-aware approach.

\section{Future Work}

A natural continuation of this work is to include a greater range of machine learning models in our contract suite.
This will enable a large-scale empirical evaluation to evaluate the utility of our solution, and empirically identify the prevalence of semantic faults in machine learning code.
Such a study will need to account for unknown models and preprocessing steps, as new machine learning models are released frequently, which is not currently supported by our tool.
For that purpose, we will aim to exploit three-valued logic to maximize the utility for machine learning developers, while minimizing the number of incorrectly identified faults.

\bibliographystyle{IEEEtranN}
\bibliography{references-doi2bib}

\end{document}